# An Important Structural Requirement for the Superconductor Material: A Hypothesis

**Bao Ting Zhu**

School of Science and Engineering and School of Medicine, The Chinese University of Hong Kong, Shenzhen 518172, China (E-Mail: **BTZhu@CUHK.edu.cn**)

**Abstract**

On a microscopic scale, resistivity during electric conduction is caused by collisions of the free conduction electrons with the obstructing atoms or molecules of the conductor material, resulting in heat production. Based on this fundamental understanding, a hypothesis concerning a physical requirement of the superconductor material is proposed, which suggests that for superconductivity (*i.e.*, with zero resistivity) to occur, the conductor material must have nano-sized, continuous and straight vacuum tunnels inside with effective radius size large enough to allow collision-free conduction of free electrons. Besides, some of the composite atoms of the conductor should be able to readily release electrons to form the conduction band; in fact, this basic requirement is for all forms of electrical conductors, not just for superconductors. The proposed hypothesis is supported by experimental observations in the literature, and also offers a plausible explanation for some of the poorly-understood experimental phenomena observed in the past. In addition, the hypothesis offers practical strategies for the rational design of electrical conductors with (quasi-)superconductivity. Lastly, the proposed new hypothesis also suggests a novel mechanism for neural microtubule-mediated electrical quasi-superconductance in the nervous system.

## 1. Introduction

Superconductivity is a unique phenomenon discovered by Heike Kamerlingh Onnes in 1911 by demonstrating the disappearance of resistivity in mercury at temperatures below 4 K [1]. Superconductivity research has attracted enormous interest in the past several decades as advances in this technology are expected to have vast scientific and utility potentials. The widely-accepted superconductivity theory, *i.e.*, the Bardeen-Cooper-Schrieffer (BCS) theory [2], posits that the spin-paired pairs of conduction electrons (commonly referred to as the Cooper pairs) are formed in the metal conduction band. One of the electrons in the Cooper pair may electrically distort the molecular structure of the superconducting material as it moves through, creating nearby a short-lived concentration of positive charge. The other electron in the pair may then be attracted toward this positive charge. It has been suggested that such coordination between electrons can prevent them from colliding with composite molecules of the conductor and thus eliminate electrical resistance. Based on this theory, when temperature decreases, the number of Cooper pairs would, in principle, increase, and ultimately the superconductor becomes a fully spin-paired, diamagnetic system. The theory offered a good explanation for the lower temperature superconductors, but it did not appear to offer a fully-satisfactory explanation for some of the other unsolved problems, such as the physical meaning of the superconducting critical temperature $T_c$, the mechanism of high-temperature superconductors, and the superconductivity of carbon-based nanotubes or graphene sheets. In this communication, a hypothesis concerning an important physical requirement of the superconducting material is proposed, and the inclusion of this hypothetical element may help better explain the phenomenon of superconductivity on the basis of the existing BCS theory.

## 2. A hypothesis

On a microscopic scale, it is accepted that resistivity is caused by collisions of the conduction electrons with the obstructing atoms or molecules in the conductor material, which results in heat production and increase in conductor temperature. Based on this fundamental understanding, it is hypothesized that for superconductivity to occur, the conductor material should meet the following two physical requirements:

**Requirement 1.** As depicted in **Figure 1a**, the conductor material must have nano-sized, continuous and straight vacuum tunnels inside with radius size large enough to enable collision-free conduction of free electrons, *i.e.*, zero resistivity. A number of factors, such as the lattice structure and its uniformity, the atomic size and composition, temperature and external pressure, would all affect the real size (*i.e.*, the effective radius) of the straight conduction tunnels which would then affect superconductivity. The influence of these as well as other factors on the real effective size of the vacuum conduction tunnels and superconductivity is discussed in *section* **3**.

**Requirement 2.** The composite atoms of the conductor need to be able to facilely release electrons to form the conduction band, which often is achieved through incorporation of certain metallic atoms [3], although certain non-metallic molecules can also readily release electrons for this purpose (such as in the case of neural microtubules as discussed later). This requirement is not special for superconducting materials, as it is also the case for all other forms of electrical conductors. If the conductor material cannot facilely release electrons to form the conduction band, then superconductivity would be difficult or impossible to achieve even when straight and continuous vacuum tunnels with suitable effective radius size are present inside the conductor material for collision-free conduction of free electrons.

## 3. Evaluation of the hypothesis and the supporting evidence

As briefly discussed below, the proposed hypothesis agrees well with many well-known key experimental observations described in the literature, and can also be used to offer a plausible explanation for some poorly-understood experimental phenomena observed in the past.

### 3.1. The physical meaning of the superconducting critical temperature $T_c$

It is known that superconductivity often is more readily observed at ultra-low temperatures (near 0 K). The reason for this phenomenon is because at ultra-low temperatures, the atomic movements (such as agitation and vibration) of the conductor's composite atoms will be close to zero, *i.e.*, the atomic movements are literally static. Here let us simply assume for a moment that there are indeed nano-sized, continuous and straight vacuum tunnels which exist inside the lattice structures of a given conductor at near 0 K when the composite atoms are almost completely static, and that the effective radius size of these conduction tunnels is large enough to allow collision-free conduction of free electrons (as depicted in **Figure 1a**). Under such conditions, the degree of atomic movements, which increase with the increase of temperature, would gradually reduce the effective passage size (*i.e.*, the effective radius) of the conduction tunnels and eventually begin to partially or even fully obstruct the existing tunnels for the collision-free conduction of free electrons (**Figure 1b**). The degree of obstruction would depend on two main factors: the initial size (*i.e.*, the effective radius) of the straight and continuous passage tunnels at temperature near 0 K, and the magnitude of the atomic movements of the conductor's composite material at a given temperature above 0 K. While the former is a built-in structural characteristic of the conductor material, which is determined by its atomic composition, the unique lattice structure and its uniformity (rigidity), the latter (*i.e.*, the magnitude of atomic movements) would be jointly determined by the internal atomic/molecular bonding structures of the conductor, temperature, external pressure, and others.

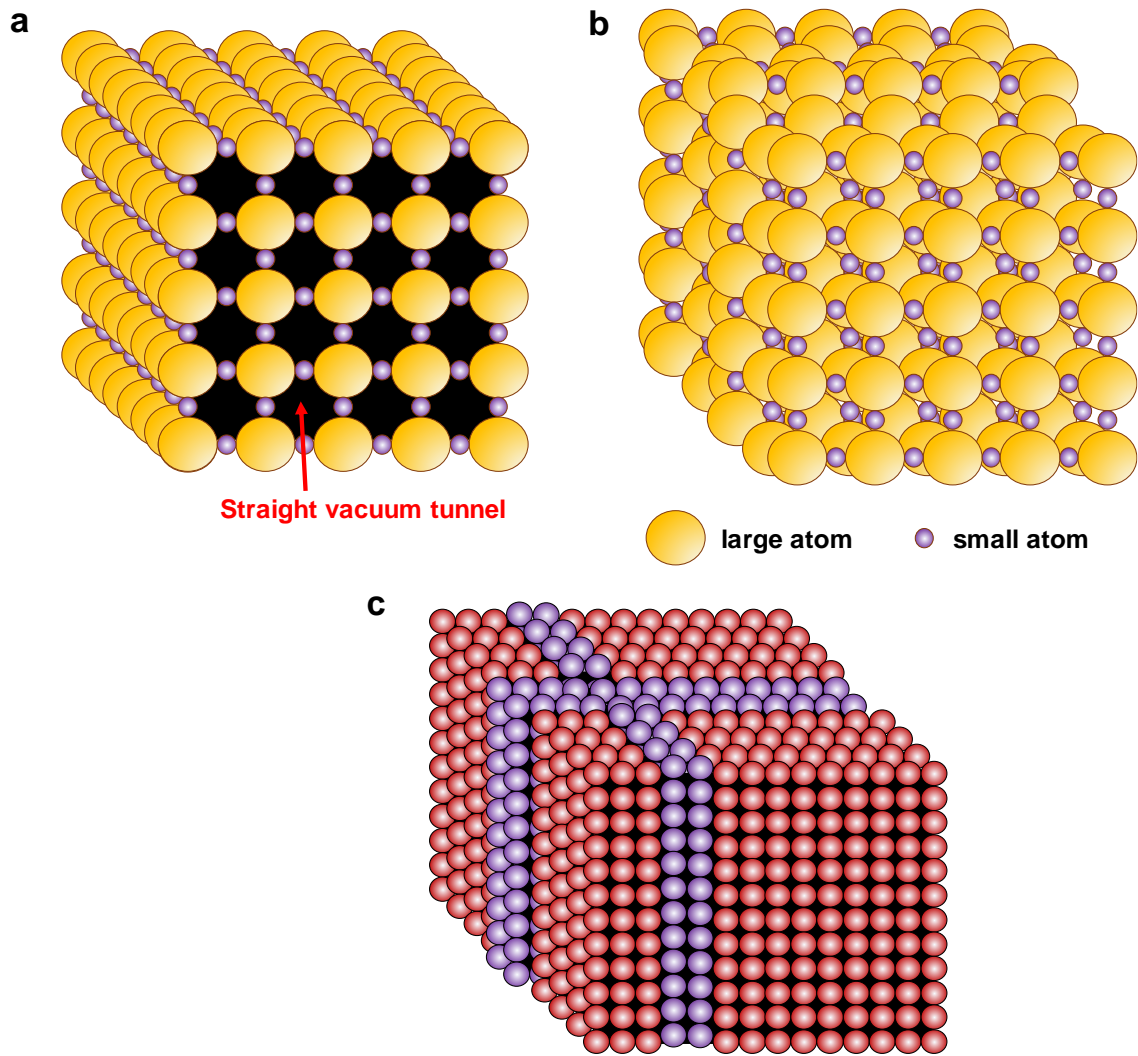

**Figure 1. A schematic depiction of the hypothetical lattice structures made of atoms with different sizes. a, b.** A hypothetical conductor made of a group of large atoms and a group of small atoms stacking together according to certain stacking rules. At near 0 K, the atoms are nearly static (with little or no vibrational movements), and the lattice structure leaves large, straight and continuous vacuum tunnels inside (*i.e.*, the black regions in **a**) that can allow collision-free conduction of free electrons. However, when the temperature increases above $T_c$, increased atomic movements would diminish the effective radius size and eventually abolish the straight and continuous passage tunnels inside and also its superconductivity (as depicted in **b**). **c.** A hypothetical conductor made of a single type of small metal atoms (such as copper) which contains small imperfections in its internal lattice structure (labelled with purple color). Because of the small atomic radius which can only form very small vacuum tunnels inside, even small imperfections in the lattice structure (labelled with purple color) may completely abolish their chances to form straight and continuous tunnels inside with sufficiently-large effective radius size to allow collision-free conduction of free electrons (even at very low temperatures when all composite atoms are nearly static).

If we assume that at or above a certain temperature when the atomic movements of the conductor's composite material will reach a degree that begins to diminish the effective radius size of the straight and continuous conduction tunnels and thereby abolish the collision-free conduction of free electrons, then this temperature would theoretically be very close to what is commonly referred to as the superconducting critical temperature, $T_c$, for a given superconductor. Stated differently, at or near $T_c$, the effective size of the straight and continuous vacuum tunnels would have the minimally-required effective radius ($R_m$) for collision-free conduction of free electrons. At temperatures below $T_c$, the degree of atomic movements in the conductor material will be smaller, which would mean that the effective radius size of the straight vacuum conduction tunnels might become larger than the minimally-required radius $R_m$ for the collision-free conduction of free electrons, thereby facilitating superconductivity at temperatures below $T_c$. However, at temperatures above $T_c$, the higher degrees of atomic movements of the conductor's composite material likely would cause a narrowing of the effective radius size of the electron passage tunnels, thereby reducing or completely abolishing superconductivity.

It is of note that at or near $T_c$ for a given superconductor, the minimally-required radius $R_m$ of the straight and continuous vacuum tunnels required for collision-free conduction of free electrons is much smaller than the estimated de Broglie wavelength ($\lambda$) of the moving electrons ($\lambda = h/p$, where $h$ is the Planck constant and $p$ is an electron's momentum). According to the classical approximations for momentum $p$ ($p = mv$, where $m$ is the mass of an electron and $v$ is its conduction speed) and kinetic energy $K$ ($K = ½ mv^2$), the wavelength of a moving electron can be estimated according to its kinetic energy. For a conduction election with a kinetic energy $K$ in the range of 1–10 eV, its estimated wavelength would be 1.23–0.39 nm. However, even when $\lambda$ = 0.39 nm, it is apparent that this wavelength is still far greater than the radius of the straight vacuum tunnels existed inside most superconducting materials. Based on this information, it is apparent that the electrons of the conduction band move inside the superconducting materials in a manner that is independent of its de Broglie wavelength ($\lambda$). It is assumed that the free conduction electrons in the conductor (such as a metal) would move throughout the vacuum space inside the conductor as free electrons. In addition, it is assumed that the electrons do not collide with one another but only with the atoms of the conductor's composite material.

### 3.2. Some structural properties of the superconducting material

Understandably, some of the conductors may indeed meet the above two physical requirements to be superconductors, *i.e.*, they contain the electron-releasing atoms or molecules, and they also contain straight and continuous vacuum tunnels inside with effective radius size large enough to allow collision-free conduction of free electrons. However, the chances for the presence of straight and continuous vacuum tunnels for collision-free conduction of free electrons inside a naturally-occurring conductor likely are quite low, and this might be the main reason why many naturally-occurring conductors cannot function as superconductors under most experimental conditions.

On the other hand, it is of note that some superconductors may happen to contain larger, straight and continuous vacuum conduction tunnels inside their lattice structure, *i.e.*, their effective passage size at or near 0 K is far bigger than the minimally-required effective radius ($R_m$) for collision-free conduction of free electrons. As such, these superconductors would have higher $T_c$ because the vacuum tunnels inside have sufficiently-large effective radius size which can tolerate certain degrees of vibrational movements of the conductor's composite atoms and yet can still allow collision-free passage of free conduction electrons at temperatures well above $T_c$.

As a general principle, it is assumed that all electrical conductors (not just superconductors) contain vacuum tunnels inside for the conduction of free electrons. However, when the vacuum tunnels inside a conductor are discontinuous and not straight, the conduction of free electrons inside these vacuum tunnels would always meet with resistance as they would constantly collide with the obstructing atoms of the conductor's composite material. Based on this understanding, it is crucial that for a conductor to become a superconductor, there must be straight and continuous vacuum passage tunnels inside that will enable collision-free conduction of free electrons. To meet this requirement, it is apparent that the following two factors would, in some ways, jointly affect the superconductivity of a conductor: One is the high degree of uniformity of the lattice structure of the conductor material which would especially favor the formation of straight and continuous vacuum tunnels inside for collision-free conduction of free electrons; the other factor is the atomic size, as larger atoms, in general, would be more favorable for the formation of larger vacuum tunnels inside. As discussed below, this general concept is in line with some of the experimental observations described in the literature:

*i*. Data in **Table 1** is a brief comparison of the relationship between the superconductivity of several pure metals and their atomic numbers, atomic radii, and hardness indices. Here the Mohs hardness index (MPa) of a conductor material is used as a gross parameter to partially reflect the degree of rigidity and uniformity of the metallic atoms contained inside a conductor. It is assumed that when the hardness of a conductor material is higher, its structural rigidity and uniformity likely would also be higher (although this is not always true). Interestingly, when the hardness indices of the metal conductors and their atomic sizes are jointly considered, it becomes surprisingly apparent that those pure metals with higher degrees of hardness plus relatively large atomic sizes (such as niobium, tantalum, iridium, vanadium and tungsten) all are more apt to function as superconductors under certain conditions. This information is quite intriguing and offers partial support for the importance of atomic size and, in particular, the degree of uniformity of the lattice structure in affecting the superconductivity of a metal conductor. However, some may argue that the case of mercury (Hg) does not seem to be in agreement with this general concept. Here it should be noted that while Hg is in a liquid form at room temperature, it is possible that the solid Hg at near 0 K may form a highly-organized and uniform lattice structure that would enable the formation of straight and continuous vacuum tunnels inside for collision-free

**Table 1. Relationship between the superconductivity of several pure metal Conductors and their atomic numbers, atomic radii and Mohs hardness.**

| Element (symbol) | Atomic number | Atomic radius (Å)[1] | Superconductivity[2] | Mohs hardness (MPa)[3] |
|---|---|---|---|---|
| Mercury (Hg) | 80 | 1.62 | Yes ($T_c$ = 4.153 K) | – |
| Barium (Ba) | 56 | 2.24 | No | 1.25 |
| Lead (Pb) | 82 | 1.75 | Yes ($T_c$ = 7.193 K) | 1.5 |
| Silver (Ag) | 47 | 1.44 | No | 2.5 |
| Gold (Au) | 79 | 1.44 | No | 2.5 |
| Aluminum (Al) | 13 | 1.43 | Yes ($T_c$ = 1.196 K) | 2.75 |
| Copper (Cu) | 29 | 1.28 | No | 3 |
| Iron (Fe) | 26 | 1.27 | No | 4 |
| Cobalt (Co) | 27 | 1.26 | No | 5 |
| Niobium (Nb) | 41 | 1.48 | Yes ($T_c$ = 9.25 K) | 6 |
| Tantalum (Ta) | 73 | 1.48 | Yes ($T_c$ = 4.483 K) | 6.5 |
| Iridium (Ir) | 77 | 1.36 | Yes ($T_c$ = 0.14 K) | 6.5 |
| Vanadium (V) | 23 | 1.35 | Yes ($T_c$ = 5.3 K) | 7 |
| Tungsten (W) | 74 | 1.41 | Yes ($T_c$ = 0.012 K) | 7.5 |

***Notes:***

[1] The atomic radius information is obtained from the web-link: baidu.com, accessed on February 1, 2023.

[2] The superconductivity information is obtained from the web-link: chazidian.com, accessed on February 1, 2023.

[3] The Mohs hardness information is obtained from the web-link: Mohs Hardness of All the Elements in the Periodic Table, accessed on February 1, 2023.

conduction of free electrons, thereby making it a superconductor at near 0 K.

*ii.* It is of note that the metallic atom barium (Ba) has been among the most widely-used elements for making superconducting materials (reviewed in [3]). The higher rate of past success with Ba likely is because this element has a relatively large atomic radius. When it is used in combination with other smaller metallic or non-metallic atoms (as schematically depicted in **Figure 1a**), it has a higher chance of forming suitable lattice structures that may leave straight conduction tunnels inside with larger effective radius size, thus favoring the collision-free conduction of free electrons (as schematically depicted in **Figure 1a**). To put it figuratively, if one tries to stack a large number of different-sized balls together according to certain repetitive stacking rules, the chances of successfully forming larger straight passage tunnels within this stack of balls would be greater if one tries to stack a proper ratio of large-size balls (*e.g.*, basket balls) and smaller balls (*e.g.*, tennis balls; the small balls may function as structure-stabilizing bonding elements or as facile electron releasers) together, as opposed to stacking all small balls together (*e.g.*, tennis balls or ping-pong balls; either in certain mixture or just a single type of small balls). In this context, it is worth mentioning that the metallic element cesium (Cs), which has an even larger atomic radius than Ba, theoretically would be an excellent choice for making superconducting materials [6]. However, this element is chemically not as stable compared to Ba, which might have limited its practical use in many instances.

In reality, many of the well-known experimentally-prepared superconductor materials were usually made of a combination of large and small atoms that have their crystal lattice structures favoring the formation of larger passage tunnels inside. For instance, the classic perovskite oxide $SrTiO_3$ [5] made in 1964 consists of two large-size atoms and one small oxygen, which leaves considerable tunnel space in its crystal lattice structure. Similarly, several other well-known superconducting materials, such as $YBa_2Cu_3O_7$, $La_2CuO_4$ and $MgB_2$, all appear to have unique crystal lattice structures that favor the formation of sizable passage tunnels within (discussed in [3]). Certainly, it will be of considerable interest in the future to model and also experimentally determine if there are straight and continuous passage tunnels contained in these superconductors.

*iii*. An example that is somewhat opposite to the above example of barium (Ba) is the case of pure copper (Cu), which is an excellent regular conductor but cannot become superconducting even at very low temperatures. Because the copper atoms have a very small atomic radius which would only form very small vacuum tunnels inside, even very small imperfections in the lattice structures might drastically reduce or completely abolish the chances to form straight and continuous tunnel spaces inside with sufficiently-large effective radius size that can allow collision-free conduction of free electrons (as schematically depicted in **Figure 1c**). Probably because of these reasons, Cu cannot effectively serve as a superconductor under almost all experimental conditions.

*iv*. Here it is also relevant to note that the free hydrogen atoms, which are the smallest atoms with a facile tendency to release their single electrons, when tightly stacked together with relatively larger atoms through chemical bonding and/or under super-high pressures, may favor to produce suitable lattice structures that would allow collision-free conduction of electrons within the conductor. The hydrogen atom is suitable for this type of experimental design likely because after its single electron is released to form the chemical bond or join the conduction band, the remaining proton will be much smaller than the hydrogen atom, thereby potentially leaving adequate space for collision-free conduction of free electrons. This intriguing possibility appears to be in line with a number of recent experimental observations (discussed in [7, 8]).

### 3.3. Effect of high pressure on superconductivity

In the past decade or so, there have been a number of studies demonstrating that under conditions of super-high external pressure, superconductivity could be more readily observed, even at higher temperatures [9–12]. Based on the hypothesis proposed here, it is understood that the super-high external pressure can help establish superconductivity because it can significantly limit the vibrational movements of the conductor's composite atoms. Therefore, the high pressure likely would have a similar actual effect as the ultra-low temperatures, and both could help reduce the vibrational movements of the composite atoms that form the straight electron passage tunnels, thereby facilitating the establishment of superconductivity.

In this context, it is of note that inside the earth's core, there is a buildup of enormous heat and pressure, due to the huge mass of the earth material lying on top of it. Although the high temperature would greatly increase the tendency of atomic movements, these movements likely are suppressed by the overwhelming high pressure exerted toward the earth's core by the mass of the earth. As such, some of the superconducting materials formed naturally inside the earth's core would be favored to exhibit superconductivity under the favorable high pressure conditions. In addition, the high temperatures at the earth's core would facilitate the release of electrons to form the conduction band. Jointly, these unique conditions at the earth's core would favor the formation of superconductivity, which might explain why the earth is a strong and giant magnet.

### 3.4. Carbon nano-devices as (quasi-)superconductors

Graphite is a layered material made of carbon sheets, and each layer of the carbon atoms was tightly packed in a two-dimensional honeycomb crystal lattice, called graphene [13]. Carbon nanotubes, discovered by Iijima in 1991 [14], have scroll-type structures (4–30 nm in diameter) and are commonly referred to as multi-walled nanotubes. In comparison, the single-wall nanotubes usually have a size of approximately 1 nm in diameter. Earlier studies have surprisingly found that carbon nonotubes and graphene sheets are good conductors [15, 16], whereas diamond, which is also made of carbon, is a good insulator. In light of the proposed new hypothesis, it is understood that in the case of carbon nanotubes, the conduction electrons released inside by doped elements or introduced into these straight vacuum nanotubes would be able to move essentially collision-free, which is similar to the conduction electrons moving collision-free through the straight and continuous vacuum tunnels inside a metallic superconductor. In a similar manner, it is expected that the conduction electrons may also be able to move essentially collision-free through the suitable vacuum spaces present between the two adjacent graphene sheets.

It is known that with regular metal conductors, temperature increase is mostly associated with increased resistivity. However, in most cases of carbon nanotubes or graphene sheets, resistivity decreases with increasing temperature [13]. This intriguing discrepancy can be readily explained on the basis of the new hypothesis. In the case of the normal metal conductors, it is understood that temperature increase would increase the movements of metallic atoms, which would then reduce the effective radius size of vacuum tunnels inside and thus increase collisions of the composite metal atoms with the free conduction electrons, resulting in increased resistivity and heat production. By contrast, in the case of carbon nanotubes and graphene sheets, the size (or space) of the vacuum passages for free electrons are usually far larger than those formed inside normal metallic conductors or superconductors, and as such, an increase in temperature (which is associated with increased movements of composite atoms that form the nanotubes or graphene sheets) may not cause a drastic narrowing of the effective passage size of the nanotubes or of the effective vacuum space between the two adjacent graphene sheets in such a manner that would abolish superconductivity. On the other hand, increased temperature would facilitate the release of free electrons to form the conduction band, which would actually increase conductivity. Therefore, the net effect of temperature increase in the cases of carbon nanotubes or graphene sheets often is associated with increased conductivity.

Similar to carbon nanotubes and graphene sheets, superconductivity has also been observed in a number of fullerene-based materials, including the well-known C60 Buckminster balls doped with certain metallic elements [17]. The fact that fullerene-based materials, along with carbon nanotubes and graphene, which are all carbon-based structures containing ample nano-sized vacuum spaces inside and are superconductors, clearly suggests that the presence of suitable vacuum spaces (tunnels) in the structure for collision-free conduction of free electrons is a crucial physical requirement for superconductivity.

In 2007, Novoselov *et al.* [18] discovered the room-temperature quantum Hall effect in graphene. It was suggested that the observation was due to the highly unusual nature of charge carriers in graphene, which might behave as massless relativistic particles (Dirac fermions) and move with little scattering under ambient conditions [19, 20]. In light of the proposed hypothesis, the observation appears to clearly indicate that the conduction electrons can move in a ballistic manner and mostly collision-free within the two adjacent sheets of graphene.

### 3.5. Neural microtubules as a physiological quasi-superconductor

It is known that the nervous system (brain in particular) conducts enormous amount of neuro-electrical activity almost all the time, even during sleep. Unlike man-made electronic devices (such as computers), the nervous system surprisingly never suffers from "overheating" due to its overwhelming amount of neuro-electrical transmission. Based on the new mechanistic explanation on (quasi-)superconductivity tendered above, it is hypothesized that the neural microtubules (neuro-MTs) which are major internal structural components of axons and dendrites may function as unique nano-sized bio-devices that can mediate electrical transmission with quasi-superconductivity. Neuro-MTs are hollow cylindrical tubes (with an outer and inner diameter of ~25 and 14 nm, respectively) [21], consisting of 13–15 rows of filaments [24, 25] made of $\alpha$- and $\beta$-tubulin heterodimers [22, 23] (depicted in **Figure 2a**). Notably, there are two types of microtubule lattice structure [21]: one type has the rotational symmetry, and the whole tube is continuous without a seam in which $\alpha$-subunits are always next to $\alpha$-subunits ($\alpha$–$\alpha$) and $\beta$-subunits next to $\beta$-subunits ($\beta$–$\beta$) (**Figure 2a**). Another lattice structure has a physical discontinuity known as a seam, where the $\alpha$-subunits in a protofilament associate laterally with $\beta$-subunits in the adjacent protofilament ($\alpha$–$\beta$). It is generally thought that the presence of a seam is a weak physical point in the MT structure.

It is hypothesized that the neuro-MTs contained inside axons, dendrites and nerve fibers are symmetric cylindrical nanotubes without a seam, such that they can better maintain their strong vacuum hollow structure inside suitable for collision-free conduction of free electrons during neuro-electrical transmission (**Figure 2a**).

The $\alpha/\beta$-tubulin heterodimer is an electric dipole, with a high negative electric charge of ~23e and a large intrinsic high dipole moment [26, 27]. Like other dipoles, MTs in solution can align with applied electric fields [28]. It is of note that because of this unique dipole arrangement, the inner surface of a neuro-MT actually contains consecutive positively-charged and negatively-charged cylindrical structures (discussed in more detail later).

It is known that the inner part of an axon (*i.e.*, its cytosolic compartment) is negatively charged (approximately –70 mV). The outer surface of neuro-MTs is also negatively charged as the *C*-terminal tails of $\alpha$- and $\beta$-tubulins contain several acidic residues which are located on the outer surface (reviewed in [29]). The inner surface of a neuro-MT is even more negative, as it can serve as a high-capacity electron-storage device [30, 31]. Neural stimulation in the form of action potentials occurring at neural cell membrane will activate Na$^+$ channels, resulting in Na$^+$ influx. An increase in cytoplasmic Na$^+$ at or near a neuro-MT resulting from an action potential would be

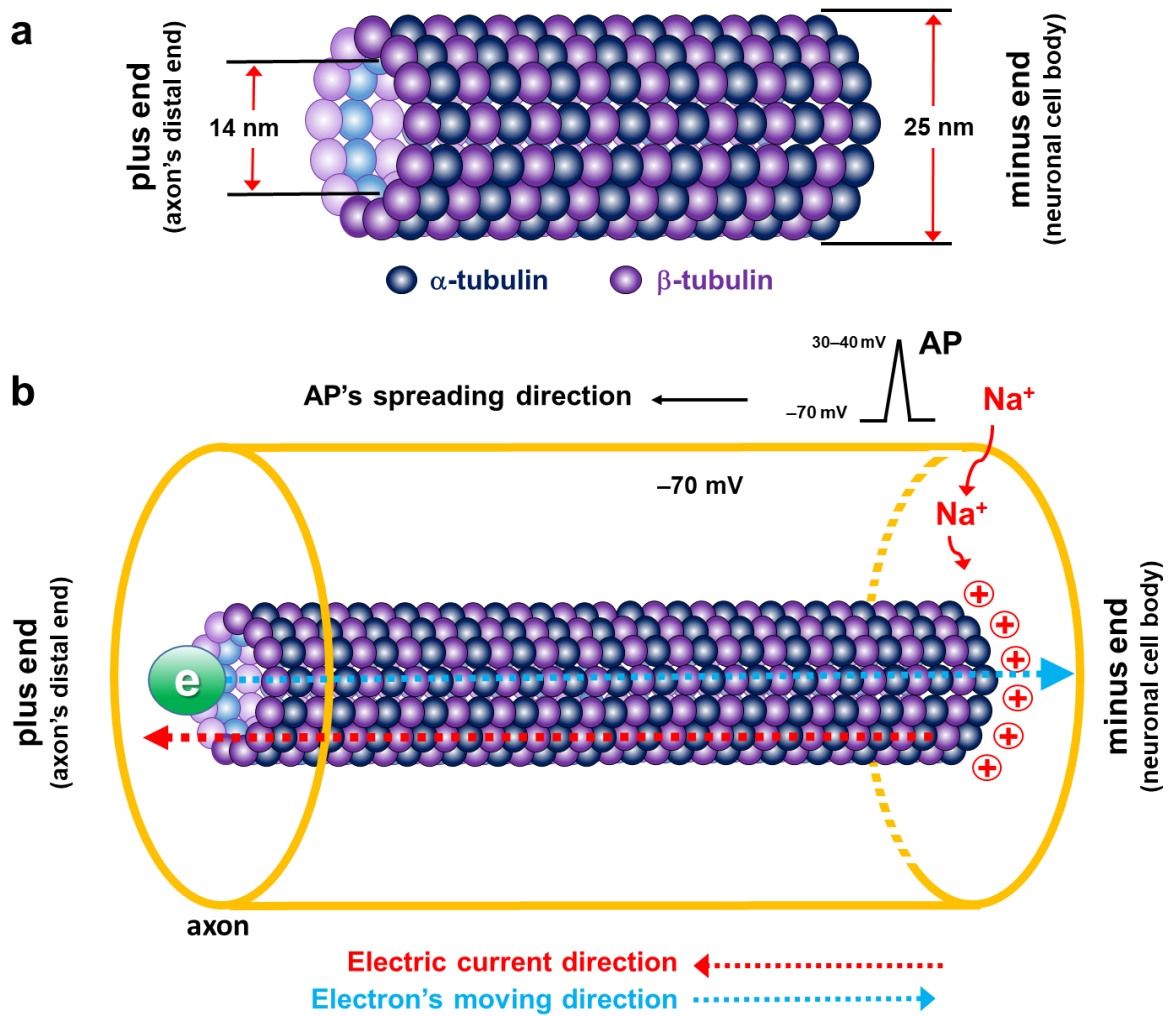

**Figure 2. A neuro-MT inside an axon may mediate neuro-electrical transmission with a unique form of quasi-superconductivity. a.** A small segment of a neuro-MT showing the basic symmetrical cylindrical structure which contains 13–15 filaments made of α-/β-tubulin heterodimers. Note that the neuro-MT has a continuous ring structure without a visible seam. Each α-/β-tubulin heterodimer unit can form a strong dipole. As indicated, for a neuro-MT inside an axon, it has a clear polarity, with the plus end located at the distal end of the axon and its minus end close to the cell body. This polarity of a neuro-MT is exactly opposite in a dendrite. **b.** It is hypothesized that the neuro-MT in an axon has a hollow and vacuum passage inside and can allow collision-free conduction of free electrons. When an action potential (AP) is initiated, Na+ channels at or near the trigger zone (which is the beginning part of the axon) are activated, resulting in Na+ influx accompanied by a sharp rise in voltage inside that particular segment of the axon and neuro-MT. As a result, free electrons will be released inside the neuro-MT to form the conduction band that will move toward the initiating end of a neuro-MT where it has a higher electric potential.

functionally similar to directly applying an electric field (or voltage) to a neuro-MT, triggering the release of electrons stored inside the neuro-MT, and the free electrons will move toward where the AP is initiated (**Figure 2b**). It is hypothesized that the conduction electrons formed inside the vacuum neuro-MT would have the following characteristics:

*i.* The free conduction electrons will be moving in a ballistic manner in the center of the hallow MT, due to the circular forces exerted evenly on the conduction electrons by the consecutive cylindrical dipoles (made of $\alpha/\beta$-tubulin heterodimers). Because neuro-

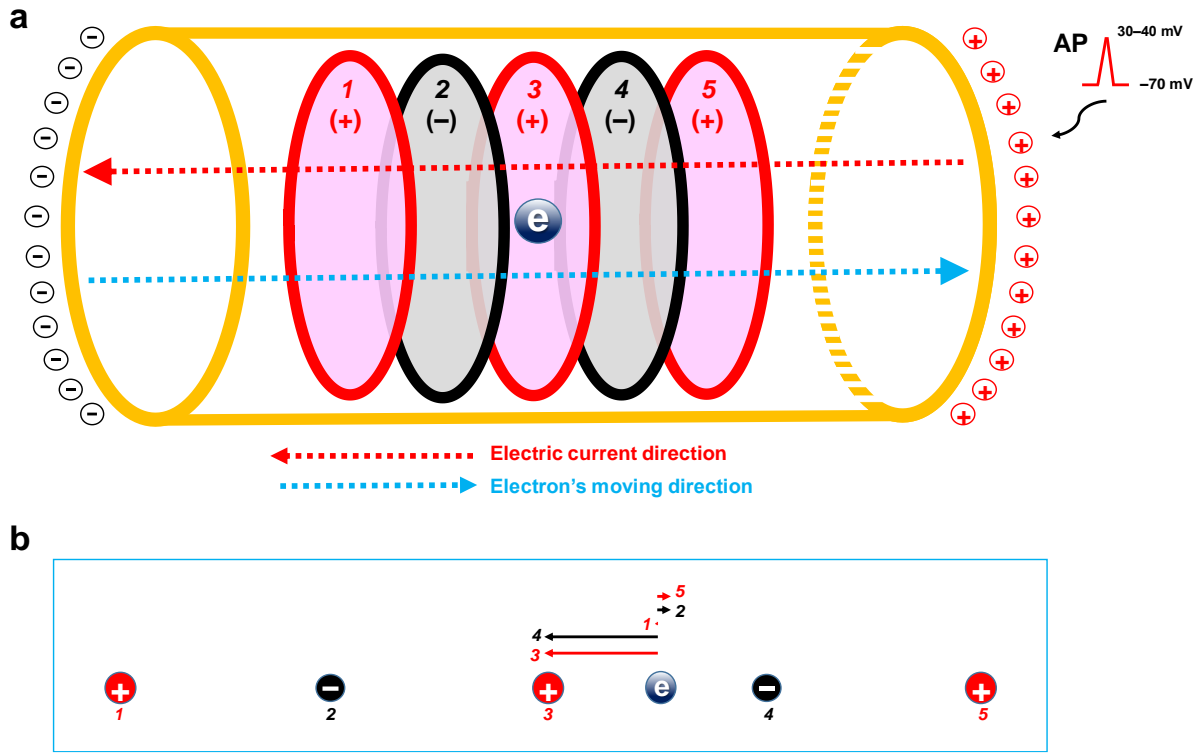

**Figure 3. Effect of the consecutive dipole-ring structures on the conduction of free electrons inside a neuro-MT.** As depicted in **a**, when a conduction electron is right at the center of the positively-charged ring *3*, all the forces generated by the neighboring dipole rings *1–5* that act on the electron will cancel each other out. However, when the electron is right in the middle of the positively-charged ring *3* and the negatively-charged ring *4* as schematically depicted in **b**, the forces generated by the neighboring rings will strongly slow down its momentum moving through a positively-charged ring (in this case, ring *3*) to the next positively-charged ring (*i.e.*, ring *5*). Note that the length of each arrow is drawn proportional to the exact magnitude of the calculated force. In addition, as soon as the neural stimulation (*i.e.*, the AP or voltage) disappears, the free electron conduction is expected to come to a complete stop almost instantaneously, and the free conduction electrons will be forced to stop exactly at the locations where the positively-charged rings are.

MTs are expected to be vacuum in nature, the conduction of free electrons inside the MTs would be collision-free (*i.e.*, superconductivity) (**Figure 2b**).

*ii.* Owing to the presence of consecutive dipole ring structures of the neuro-MTs (as depicted in **Figure 3a**), the speed of the conduction electrons in each neuro-MT is expected to be far slower than usual, as the two neighboring rings (one positively-charged and one negatively-charged) will each exert a force on the conduction electrons and slow down their moving through each positively-charged ring. Despite the slow conduction speed, it should be noted that the speed by which an electric current is established throughout the entire length of a neuro-MT theoretically would not be affected at all. This unique feature will enable physiological neuro-electrical transmission to occur with super-high efficiency as only minimal numbers of the free conduction electrons will actually pass through a neuro-MT within unit time. Even if we assume that the kinetic energy ***K*** of a slow free conduction electron is as low as 0.1 eV, its estimated wavelength (**λ**) would be 3.88 nm, which is still smaller than the inner radius (~7 nm) of a neuro-MT. This is probably why neuro-MTs have a relatively large inner radius size, as it would enable the conduction of electrons with very low speed inside neuro-MTs and thereby would help conserve cellular energies (**Figure 3a**).

*iii.* As explained in **Figure 3b**, the presence of the consecutive cylindrical dipole structures inside a MT also helps to terminate the conduction band with exceptional high efficiency, *i.e.*, as soon as the neural stimulation (in the form of an action potential) ends. This unique feature is fully in line with the known characteristics of physiological neuro-electrical activities (for instance, the voluntary muscle movements can be initiated at will almost instantaneously, and they are also terminated immediately without any lingering after-effects). Because of this unique feature, the collision-free electric conduction occurring inside a neuro-MT is considered a unique form of quasi-superconductivity.

*iv.* As aforementioned, neuro-MTs can serve as a high-capacity charge storage device [30, 31]. The large capacitance of neuro-MTs for electrons is a crucial feature which would enable them to fulfill the vital physiological function of mediating enormous amount of neuro-electrical transmissions on a constant basis.

## 4.  Concluding remarks

It is hypothesized that for (quasi-)superconductance to take place, the conductor must have straight and continuous vacuum tunnels inside with effective radius size large enough to allow essentially collision-free conduction of free electrons. The proposed hypothesis is in agreement with many experimental observations, and also offers a plausible explanation for some poorly-understood experimental phenomena of the past. Based on the proposed hypothesis, it is further suggested that the neuro-MTs contained inside axons, dendrites and nerve fibers may serve to mediate neuro-electrical transmission with a unique form of quasi-superconductivity.

It is of note that the proposed hypothesis also offers some practical suggestions for the rational design of (quasi-)superconductors for research and industrial uses. For the more traditional element-based superconductors, it might be possible in the future to rationally design (or virtually screen for) unique three-dimensional metallic lattice structures that may contain suitable vacuum tunnels inside for collision-free electric superconductance. On the other hand, in the cases of carbon nanotubes or graphene sheets, future efforts are needed to improve the fabrication techniques to ensure the production of more uniform vacuum nanotubes with sizes properly suited for electrical superconductance. Similarly, it is expected that future conducting devices based on graphene sheets will also be of similar values, but technological advances are needed to obtain uniformity and optimal thickness (and size) of the vacuum spaces between the two adjacent sheets. It is expected that when multi-layered graphene sheets (with optimal vacuum spacing) are evenly stacked up, or multiple graphene sheets are rolled up to form multi-walled nanotubes, superconductance (or quasi-superconductance) may occur inside the vacuum spaces of the multi-layered structures. Lastly, it is of note that not all perfectly-fabricated carbon nano-conducting devices can serve as good conductors. The ability to firmly and evenly incorporate charge-producing composite elements inside the vacuum nanotubes or between the graphene sheets is another challenge that needs to be overcome before carbon-based nano-devices can reach their full potentials for future use as (quasi-)superconducting devices. Similarly, it is expected that other atomic elements (*e.g.*, silicone) will also be widely used in the future in place of carbon in similar nano-devices with (quasi-)superconductivity.

**Acknowledgements**

The author wishes to thank Prof. Yan Zhou and Prof. Xi Zhu at The Chinese University of Hong Kong, Shenzhen, for discussion of the kinetic energy levels of the conduction electrons in a regular metal conductor. The author also wishes to thank Mr. Yong-Xiao Yang, a researcher in the author's laboratory, for helping compile the raw information used in Table 1. An initial version of this paper, entitled, "Mechanism of Superconductivity: A Theory" was first posted on arXiv (2207.01226) on July 4, 2022.

**Conflict of Interest**

The author declares no conflict of interest.